\documentclass[12pt]{article}

\usepackage{fullpage}
\usepackage{amssymb}
\usepackage{bbm}
\usepackage{color}
\usepackage{ulem}

\newcommand{\Comment}[1]{{}}
\definecolor{Blueish}{rgb}{0.27,0.17,0.75}
\usepackage[linktocpage=true]{hyperref}
\hypersetup{
colorlinks=true,
citecolor=Blueish,
linkcolor=Blueish,
urlcolor=Blueish,
pdfauthor={},
pdftitle={},
pdfsubject={}
}

\DeclareFontFamily{OT1}{rsfs10}{}
\DeclareFontShape{OT1}{rsfs10}{m}{n}{ <-> rsfs10 }{}
\DeclareMathAlphabet{\mathscript}{OT1}{rsfs10}{m}{n}

\def\gsim{ \lower .75ex \hbox{$\sim$} \llap{\raise .27ex \hbox{$>$}} }
\def\lsim{ \lower .75ex \hbox{$\sim$} \llap{\raise .27ex \hbox{$<$}} }
\def\be{\begin{equation}}
\def\ee{\end{equation}}
\def\bea{\begin{eqnarray}}
\def\eea{\end{eqnarray}}

\newcommand{\ns}{\normalsize}

\newcommand{\rd}{{\rm d}}

\usepackage{latexsym,amsmath,amssymb,epsfig}

\usepackage{graphicx}
\usepackage{graphicx,subfigure}
\usepackage{epstopdf}
\usepackage[body={17.5cm, 21.1cm},right=2cm]{geometry}
\usepackage{amssymb}
\usepackage{amsmath}
\usepackage{psfrag}
\usepackage{epsfig}
\usepackage{cancel}
 \allowdisplaybreaks[4]

\usepackage[all]{xy}

\begin{document}

\begin{titlepage}

\title{
  \hfill{\ns } 
   {\LARGE Strong Coupling Problem with Time-Varying Sound Speed}
\\[0.5em] }
\author{
   Austin Joyce and Justin Khoury
     \\[0.3em]
{\ns  Center for Particle Cosmology, Department of Physics and Astronomy,} \\[-0.2cm]
{\ns  University of Pennsylvania, Philadelphia, PA 19104}
}
\date{}

\maketitle

\begin{abstract}
For a single scalar field with unit sound speed minimally coupled to Einstein gravity, there are exactly three distinct cosmological solutions which produce a scale invariant spectrum of curvature perturbations in a dynamical attractor background, assuming vacuum initial conditions: slow-roll inflation; a slowly contracting adiabatic ekpyrotic phase, described by a rapidly-varying equation of state; and an adiabatic ekpyrotic phase on a slowly expanding background. Of these three, only inflation remains weakly coupled over a wide range of modes, while the other scenarios can produce at most 12 e-folds of scale invariant and gaussian modes. In this paper,  we investigate how allowing the speed of sound of fluctuations to evolve in time affects this classification. While in the presence of a variable sound speed there are many more scenarios which are scale invariant at the level of the two-point function, they generically suffer from strong coupling problems similar to those in the canonical case. There is, however, an exceptional case with superluminal sound speed, which suppresses non-gaussianities and somewhat alleviates strong coupling issues. 
We focus on a particular realization of this limit and show these scenarios are constrained and only able to produce at most 28 e-folds of scale invariant and gaussian perturbations. A similar bound should hold more generally --- the condition results from the combined requirements of matching the observed amplitude of curvature perturbations, demanding that the Hubble parameter remain sub-Planckian and keeping non-gaussianities under control. We therefore conclude that inflation remains the unique cosmological scenario, assuming a single degree of freedom on an attractor background, capable of producing arbitrarily many scale invariant modes while remaining weakly coupled. Alternative mechanisms must inevitably be unstable or rely on multiple degrees of freedom.
\end{abstract}

\end{titlepage}
\setcounter{page}{2}

\section{Introduction}
There is clear evidence for a nearly Harrison--Zel'dovich and gaussian spectrum of perturbations from cosmic microwave background and large scale structure observations. One of the most pressing questions in early--universe cosmology is to understand the mechanism underlying the generation of this primordial spectrum. Inflation~\cite{inf} is one explanation, but it is important to understand what other scenarios predict a spectrum of curvature perturbations consistent with observations. This question has motivated much research into alternative theories; for instance the pre--big bang scenario of \cite{Gasperini:1992em}$-$\cite{Gasperini:2007vw}, string gas cosmology \cite{Brandenberger:1988aj}$-$\cite{Battefeld:2005av} and ekpyrotic cosmology~\cite{Khoury:2001wf}$-$\cite{Lehners:2010fy}. Production of a scale invariant spectrum of fluctuations by itself is not enough --- it is also highly desirable for the background solution to be an {\it attractor}.\footnote{The requirement that the background be an attractor may not be essential. Indeed, there are scenarios where instabilities play a crucial role and have important consequences, for example the matter bounce scenario \cite{Brandenberger:2009yt} in the single--field case, as well as the curvaton mechanism and the phoenix universe \cite{Lehners:2008qe} in the multi--field case. See \cite{baumann} for a detailed discussion of single--field, non--attractor scenarios.}
Technically, this is achieved by demanding that the curvature perturbation on uniform-density hypersurfaces, $\zeta$, goes to a constant in the long wavelength limit $k\to0$. In this limit, $\zeta \approx \delta a/a$ is interpreted as a constant perturbation of the scale factor, which may therefore be absorbed locally by a spatial diffeomorphism~\cite{weinberg}.

For models involving a single, canonical scalar field ({\it i.e.} with unit sound speed, $c_s = 1$) minimally coupled to Einstein gravity, it has been shown recently that there are only {\it three} independent cosmological solutions which produce a scale invariant spectrum of curvature perturbations on an attractor background~\cite{godfrey, baumann}, assuming vacuum initial conditions. See~\cite{Piao:2003ty}$-$\cite{Kinney:2010qa} for related work. The most well--known of these solutions is of course inflation~\cite{inf}, which relies on exponential expansion of the background with $\epsilon \equiv -\dot{H}/H^2 \simeq 0$. More recently, the adiabatic ekpyrotic~\cite{adiabaticek, adiabaticek2} scenario has been proposed, in which a scale invariant spectrum is produced by a rapidly evolving equation of state $\epsilon\sim 1/t^2$ on a slowly contracting background. The third solution can be viewed as a variant of adiabatic ekpyrosis, where curvature perturbations are again sourced by a rapidly changing equation of state, but this time on a slowly expanding background~\cite{expanding}. At the level of the two-point function, these three scenarios yield indistinguishable power spectra.

However, the degeneracy is broken at the three--point level. The non--inflationary solutions have strongly scale dependent non--gaussianities~\cite{adiabaticek2, godfrey}, which can be traced to the rapid growth of the equation of state parameter. In these models, $f_{\rm NL}\sim k$ grows rapidly at small scales and perturbative control is lost when $f_{\rm NL}\zeta \sim 1$. This difficulty can be avoided by suitably modifying the potential so that $\zeta$ becomes much smaller on small scales. But this in turn restricts the range of scale invariant and gaussian modes to about 5 decades ($\sim 10^5$) or $\simeq 12$ e-folds in $k$--space \cite{adiabaticek2, godfrey}. As a result, with $c_s = 1$ and attractor background, inflation is the unique single field mechanism capable of producing many decades of scale invariant and gaussian perturbations. (Of course, as a theory of the early universe inflation must still surmount some foundational issues, such as the measure problem and low-entropy initial conditions~\cite{gibbonsturok}. 
Here we leave aside these critical questions and note that inflation, viewed as a mechanism for generating density perturbations, remains weakly--coupled over a large range of modes.)

In this paper we generalize the analysis to the case of time-varying sound speed, $c_s(t)$, as obtained, for instance, with non-canonical scalar fields. 
With Einstein gravity plus a single degree of freedom, the sound speed is the only remaining knob at our disposal.\footnote{One could also consider alternative theories
of gravity. Our analysis applies to any theory of gravity which admits an Einstein frame description in terms of some field variables, such as generic scalar tensor theories.}
As shown in~\cite{piazza}, allowing for $c_s(t)$ greatly broadens the realm of allowed cosmologies that yield a scale invariant power spectrum. In particular, any cosmology with constant equation of state
can be made scale invariant by suitably choosing the evolution of the sound speed. In this work we show
that non-gaussianities impose stringent constraints on the allowed cosmologies. Our analysis is very general and applies to {\it arbitrary} time-dependent $\epsilon(t)$ and $c_s(t)$,
with the only restriction that the null energy condition be satisfied: $\epsilon \geq 0$.

We begin by reviewing how the time--dependence of the sound speed results in an effective cosmological background for the curvature perturbation, as was first shown in~\cite{piazza}. In this effective background, which depends both on the evolution of the scale factor and the sound speed, $\zeta$ propagates at the speed of light. We derive a consistency equation that the scale factor and the sound speed must satisfy in order to have scale invariance at the two--point level. In the spirit of \cite{piazza}, given an evolution for the scale factor, solving this equation gives a suitable evolution for the sound speed for which $\zeta$ has a scale invariant two--point function on an attractor background. This shows that a time--dependent sound speed vastly increases the degeneracy at the two--point level.

As in the canonical case, this degeneracy is generically broken by the three--point function. In particular, if the three--point function is strongly scale--dependent, we generically expect the theory to become strongly coupled either in the infrared (IR) or in the ultraviolet (UV). To avoid such perturbative breakdown, we demand that certain contributions to the three--point function be scale invariant. This turns out to
be extremely restrictive: we show that slow--roll inflation is the {\it unique} cosmology with this property. Conversely, if the three--point function is not scale invariant, then non--gaussianities will increase rapidly
with scale, resulting in a finite range ($\lsim\, 10^5$ modes) of perturbations consistent with observations, as in the canonical case. This is a remarkable fact; it is extremely surprising, in light of the vast degeneracy afforded by a variable sound speed, that slow--roll inflation should be the unique possibility.

There is, na\"ively, a way to avoid our argument; since all the vertices in the cubic action are proportional to $1/c_s^2$, it may be possible to have a strongly scale dependent three--point function while remaining weakly--coupled by taking $c_s \gg 1$. There has recently been much interesting work done related to this idea. A concrete example of this limit is given by the scenario considered in~\cite{magueijo}$-$\cite{tachyacoustic}, based on earlier ideas in~\cite{vsl}$-$\cite{piao}. In these scenarios, the sound speed is assumed $\gg 1$ in the very early universe and decreases in time. It should be noted that theories with superluminal propagation cannot be UV completed by a theory with an analytic S--matrix, for example local quantum field theory or perturbative string theory~\cite{nima}. However, for the sake of generality, we remain agnostic about the nature of any UV completion of the theories we consider.

We review the particular superluminal mechanism where $\epsilon$ and $\epsilon_s$ are constant and show that such cosmologies are also tightly constrained by strong coupling considerations. 
Again in these cases, only a finite range of scale invariant modes can be generated without hitting strong coupling. 
In order to understand this constraint, recall that the $\zeta$ power spectrum is given by
\be
P_\zeta \sim \mathcal O(\epsilon)\frac{H^2}{c_sM_{\rm Pl}^2}\sim 10^{-10}~.
\ee
It is immediately clear that at early times, when $c_s \gg 1$, there is a tension between keeping $H/M_{\rm Pl} < 1$ and maintaining the observed amplitude for density perturbations. For $\epsilon\sim\mathcal O(1)$, this keeps us from taking $c_s$ to be too large, which bounds the range of scale invariant modes to about 8 decades ($\sim10^8$). This tension may be alleviated by taking $\epsilon$ to be exponentially large, but this causes the amplitude of non--gaussianities to grow. Because $c_s$ is decreasing in time, demanding that the theory remain weakly coupled imposes a bound of about 12 decades ($\sim 10^{12}$) on mode production. Though this range is significantly larger than the $\sim10^5$ viable modes in the canonical case, we still have to account for the coincidence that these overlap with the observable window.

Our results have some overlap with that of Baumann {\it et al.}~\cite{baumann}, who reached similar conclusions. However, our analysis is much more general --- in the variable sound speed case, Baumann {\it et al} considered only sub--luminal $c_s$ and nearly constant $\epsilon$. In contrast, we allow {\it arbitrary} evolution for $\epsilon$ and $c_s$ --- including the superluminal regime --- in Section \ref{threepoint} and analyze in detail the constant $\epsilon$ superluminal scenarios in Section \ref{loophole}.

\section{Scale Invariance with Variable Sound Speed}
\label{twopoint}
We begin by considering how a varying sound speed affects the classification of scale invariance at the two--point level recently undertaken in \cite{godfrey, baumann}. We make no assumptions about the underlying dynamics, only that they may be well modeled by a perfect fluid. Perturbing around a Friedmann--Robertson--Walker (FRW) background in co--moving gauge, where $\delta\rho = 0$ and $h_{ij} = a(t)e^{2\zeta}\delta_{ij}$, the quadratic action for $\zeta$ is given by~\cite{garrigamukhanov}
\be
S_{2}=M_{\rm Pl}^2\int \rd^3x\rd\tau~z^2\left[\left(\frac{\rd\zeta}{\rd\tau}\right)^2-c_s^2\left(\partial\zeta\right)^2\right]~,
\ee
where $z \equiv a\sqrt{\epsilon}/c_s$, and $\tau$ denotes conformal time, $a\rd\tau = \rd t$. This action is familiar from canonical single field models, except for the sound speed factor multiplying the spatial gradient term and appearing in the measure factor. In order to eliminate this complication, following \cite{piazza} we define the sound horizon time coordinate by $\rd y = c_s\rd\tau$.
(Note that when $c_s={\rm constant}$, the variable $y$ measures the size of the sound horizon.) Additionally, we define
\be
q\equiv \sqrt{c_s}z = \frac{a\sqrt\epsilon}{\sqrt{c_s}}~.
\ee
In terms of these new variables, the quadratic $\zeta$ action takes the familiar form
\be
S_{2}=M_{\rm Pl}^2\int\rd^3x\rd y~q^2\bigg[\zeta'^2-\left(\partial\zeta\right)^2\bigg]~,
\ee
where primes indicate derivatives with respect to $y$. The virtue of this change of variables is manifest --- $\zeta$ now propagates luminally, but in {\it effective} cosmological background
defined both by the scale factor and the sound speed.

The mode functions of the canonically normalized field, $v\equiv \sqrt{2}M_{\rm Pl}~q\cdot\zeta$, obey the familiar-looking equation of motion
\be
v_{k}''+\left(k^2-\frac{q''}{q}\right)v_k=0~.
\label{mode}
\ee
This has exactly the same form as the corresponding equation in the canonical case, but with $q$ replacing $z$, a manifestation of the curvature perturbation evolving in an effective geometry.
Assuming the usual adiabatic vacuum, it is well known that (\ref{mode}) will yield a scale invariant spectrum of perturbations provided that
\be
\frac{q''}{q} = \frac{2}{y^2}~.
\label{qeq}
\ee
Note that modes freeze out when $k\lvert y\rvert\sim1$, which corresponds to sound-horizon crossing in the constant $c_s$ case,
hence we take $-\infty<y<0$. The solution for the mode functions is then 
\be
v_k(y) = \frac{1}{\sqrt{2k}}\left(1-\frac{i}{ky}\right)e^{iky}~,
\label{vk}
\ee
which describes a scale invariant spectrum, $v_k \sim k^{-3/2}$, in the limit $y \rightarrow 0$. 

Equation (\ref{qeq}) has two solutions, $q\sim 1/(-y)$ and $q\sim y^2$, but only the former describes a background which is a dynamical attractor. To see this, note that in the long wavelength ($k\to 0$) limit we have the following expression for the power spectrum of the solution (\ref{vk})
\be
P_\zeta = \frac{1}{2\pi^2}k^3\lvert\zeta_k\rvert^2 \sim \frac{1}{q^2y^2}~,
\ee
which is indeed independent of $k$. When $q \sim 1/(-y)$, $\zeta\to{\rm constant}$ outside the horizon, indicating perturbative stability~\cite{weinberg}. The other solution $q\sim y^2$, however, implies that $\zeta$ grows outside the horizon, $\zeta\sim y^{-3}$, signaling that the background is unstable. Since we are interested in attractor backgrounds, we henceforth ignore the $q\sim y^2$ solution.
Recalling the definition of $q$, the condition for scale invariance in an attractor background may therefore be succinctly expressed as
\be
q^2=\frac{a^2\epsilon}{c_s}=\frac{\beta}{y^2}~,
\label{inv}
\ee
where $\beta$ is an arbitrary (positive) constant. It is important to note that $a$ and $\epsilon$ are not independent degrees of freedom, but are related by $\epsilon=-\dot H/H^2$. Changing time variables to $y$, this relation becomes
\be
\epsilon=\frac{\rd}{\rd t}\frac{1}{H}= \frac{c_s}{a}\left(\frac{a^2}{c_s a'}\right)'~.
\ee
Using the condition~(\ref{inv}) for scale invariance, we can rewrite this to obtain the master equation
\be
a\left(\frac{a^2}{c_s a'}\right)'=\frac{\beta}{y^2}~.
\label{diffeq}
\ee
This equation ensures a scale invariant spectrum on an attractor background. As noted in \cite{baumann}, in the case where $c_s={\rm constant}$, this equation may be recast as a particular instance of the generalized Emden--Fowler equation. 

For completeness, we review the results of \cite{godfrey,baumann}. In the case of constant sound speed (without loss of generality, we may take $c_s = 1$), there are three distinct scale invariant solutions:

\begin{itemize}

\item {\it Inflation} is a solution where the scale factor grows as $a_{\rm inf}\sim 1/(-\tau)$ and the equation of state parameter is constant $\epsilon_{\rm inf}\ll 1$ \cite{inf}. To check that this is in fact scale invariant, we note that $q^2_{\rm inf}\sim1/\tau^2$, where $y\sim\tau$ because $c_s$ is constant.

\item {\it Adiabatic Ekpyrosis} is a solution where the equation of state parameter varies rapidly, $\epsilon_{\rm ek}\sim 1/\tau^2$, while the background remains nearly static, $a_{\rm ek}\sim 1$. Again, we can check that this gives a scale invariant spectrum $q^2_{\rm ek}\sim1/\tau^2$. In fact, this corresponds to two distinct solutions, one where the background is slowly {\it contracting} \cite{adiabaticek,adiabaticek2} and one where the background is slowly {\it expanding}  \cite{expanding}. It is important to note that in these scenarios modes freeze out on sub--Hubble scales and are subsequently pushed outside the horizon
during a contracting ekpyrotic phase with constant $\epsilon \gg 1$.

\end{itemize}

Returning to the general case, given any evolution for $a$ we can find an evolution of $c_s$ that will make the spectrum of perturbations scale invariant by solving (\ref{diffeq}). Alternatively, specifying a relation between the evolution of $c_s$ and $a$ is sufficient to determine the evolution. As a result, we see that there is an enormous amount of degeneracy at the two--point level. 

An excellent illustration of this degeneracy is the case of cosmologies with constant $\epsilon$. With constant $c_s$, as reviewed above inflation is the only solution that has constant $\epsilon$.
But for more general sound speed, there is a power-law evolution for $c_s$ that yields a scale invariant spectrum for arbitrary positive values of $\epsilon$. Indeed, constancy of $\epsilon$
and $\epsilon_s \equiv\dot c_s/Hc_s$ is sufficient to deduce the scaling solutions
\be
a\sim (-y)^\frac{1}{\epsilon + \epsilon_s-1}~,~~~~~~~~~~~~~c_s\sim (-y)^\frac{\epsilon_s}{\epsilon + \epsilon_s-1}~.
\label{acspowerlaw}
\ee
Inserting these expressions into (\ref{diffeq}), we find that the solution is scale invariant for $\epsilon_s = -2\epsilon$, in agreement with~\cite{piazza}.
\section{The Cubic Action}
\label{threepoint}
Non-gaussianities offer a powerful tool for differentiating between the different cosmologies with degenerate power spectra. Since 
the precise form of the cubic action depends the underlying physics, we must choose to parameterize the microphysics in some way.
A convenient and quite general choice is to consider a non--canonical scalar field $\phi$, described by a $P(X,\phi)$ lagrangian
\be
S=\int \rd^4 x\sqrt{-g}\left[\frac{M_{\rm Pl}^2R}{2}+P(X,\phi)\right]~,
\label{action}
\ee
where $X=-\frac{1}{2}g^{\mu\nu}\partial_{\mu}\phi\partial_{\nu}\phi$. It is straightforward to show that the energy density and sound speed are related to the choice of $P$ by~\cite{garrigamukhanov}
\be
\rho=2XP_{,X}-P\;; ~~~~~~~~ c^2_s=\frac{P_{,X}}{P_{,X}+2XP_{,XX}}~.
\ee
The cubic action for $\zeta$ is derived in \cite{maldacena,seerylidsey,kachru}. Making the transformation to the sound horizon time variable $\rd y = c_s\rd\tau$,
the action takes the form, up to a field redefinition,
\begin{align}
\begin{split}
S_3=\int \rd^3x \rd y&\left[-ac_s^2\left\{\Sigma\left(1-\frac{1}{c_s^2}\right)+2\lambda\right\}\frac{\zeta'^3}{H^3}+\frac{a^2\epsilon}{c_s^3}\left(\epsilon-3+3c_s^2\right)\zeta\zeta'^2\right.\\
&
+\frac{a^2\epsilon}{c_s^3}\left(\epsilon-2\epsilon_s +1-c_s^2\right)\zeta\left(\partial\zeta\right)^2
-\frac{2a^2\epsilon^2}{c_s^3}\zeta'\partial\zeta\frac{\partial\zeta'}{\nabla^2}\\
&\left.
+\frac{a^2\epsilon}{2c_s}\left(\frac{\eta}{c_s^2}\right)'\zeta^2\zeta'
+\frac{a^2\epsilon^3}{2c_s^3}\partial\zeta\frac{\partial\zeta'}{\nabla^2}\zeta'+\frac{a^2\epsilon^3}{4c_s^3}\nabla^2\zeta\left(\frac{\partial\zeta'}{\nabla^2}\right)^2\right]~,
\label{s3pt}
\end{split}
\end{align}
where $\eta = H^{-1}{\rm d}\ln\epsilon/{\rm d}t$, and
\be
\lambda = X^2P_{,XX}+\frac{2}{3}X^3P_{,XXX}\;;~~~~~~~\Sigma=XP_{,X}+2XP_{,XX}=\frac{H^2\epsilon}{c_s^2}~.
\ee
Although we focus on a particular class of microphysical models, namely $P(X,\phi)$ theories, the model--dependence of the action is encoded only in $\lambda$. 
All other vertices in the cubic action are functions of the scale factor and the sound speed. For example, for a DBI action, the $\zeta'^3$ term in (\ref{s3pt}) vanishes identically~\cite{piazza,kachru}. Since the form of this first term will not be material to our arguments, our analysis even at the cubic level is rather general, but there may be some potential model--dependent effects from the $\zeta'^3$ vertex which we have not considered.

To estimate non--gaussianities, a useful approximation is the {\it horizon--crossing approximation}, whereby $f_{\rm NL}$ is estimated by
\be
f_{\rm NL}\sim\left.\frac{\mathcal{L}_3}{\zeta\cdot\mathcal{L}_2}\right\rvert_{k\lvert y\rvert=1}~.
\label{fnl}
\ee
Here $\mathcal L_2$ and $\mathcal L_3$ are terms in the quadratic and cubic lagrangians, respectively.
Since temporal and spatial gradients are comparable at horizon crossing ($\partial_y\sim\partial_i\sim k$),
we may trade them freely in~(\ref{fnl}). The horizon-crossing approximation generally offers a good estimate of $f_{\rm NL}$
since modes are in their ground state at early times --- when they are far inside the horizon --- and become constant outside the horizon. 
We therefore expect non-gaussianities to peak around horizon crossing.\footnote{An important exception is the adiabatic ekpyrotic solution,
where the $\epsilon^3$ contributions peak at late times, well after horizon crossing~\cite{adiabaticek2}. Although $\zeta$ goes to a constant outside the horizon,
the rapid growth of the vertex, $\epsilon^3\sim 1/t^6$, overwhelms the suppression from $\zeta$ derivatives becoming small. Thus, the horizon-crossing
approximation is a conservative estimate of non-gaussianities.}

At a classical level, perturbations are highly non-gaussian for $f_{\rm NL}\zeta \, \gsim \, 1$, corresponding to $\mathcal L_3/\mathcal L_2\,  \gsim \,1$,
and classical perturbation theory breaks down. At a quantum level, the right hand side of~(\ref{fnl}) also offers an estimate for the magnitude of loop corrections to the two-point
function~\cite{louissarah}. Thus, classical and quantum perturbation theory break down, and the theory becomes strongly coupled, whenever 
\be
\frac{\mathcal L_3}{\mathcal L_2} \sim 1~,
\ee
or $f_{\rm NL}\zeta \sim 1$. This is the same strong coupling criterion used in~\cite{baumann}.

In particular, if $f_{\rm NL}$ is strongly scale dependent, then the growth of non-gaussianities will generically lead to a breakdown of perturbation theory
either in the IR or in the UV. (As mentioned earlier, an exceptional case is the limit $c_s \gg 1$, which will be treated separately in Section~\ref{loophole}.)
This expectation is borne out by the analysis of the canonical case~\cite{adiabaticek2,expanding,godfrey}.
To avoid strong coupling, therefore, we demand that $f_{\rm NL}$ be approximately scale invariant.  

Amongst the terms in the cubic action is the vertex\footnote{This vertex is the leading contribution to non-gaussianity in the
adiabatic ekpyrotic scenarios \cite{adiabaticek,adiabaticek2, expanding, godfrey}.}
\be
S_3 \supset \int\rd^3x\rd y ~ \frac{a^2\epsilon^3}{2c_s^3}\partial\zeta\frac{\partial\zeta'}{\nabla^2}\zeta'~.
\label{S3particular}
\ee
Evaluating this vertex at horizon--crossing, we find that its non-gaussian contribution is
\be
f_{\rm NL}^{\epsilon^3} \sim \left.\frac{\frac{a^2\epsilon^3}{2c_s^3}\partial\zeta\frac{\partial\zeta'}{\nabla^2}\zeta'}{\frac{a^2\epsilon}{c_s}\zeta\cdot\zeta'^2}\right\rvert_{k \lvert y\rvert=1}
\sim \left(\frac{\epsilon}{c_s}\right)^2. 
\label{fNLeps3}
\ee
Substituting the condition~(\ref{inv}) for scale invariance at the two--point level, $a^2\epsilon/c_s \sim 1/y^2$, this reduces to
\be
f_{\rm NL}^{\epsilon^3} \sim  \frac{1}{a^4y^4}\bigg\vert_{k \lvert y\rvert=1} ~.
\ee
Now, in order for the full three--point function to be scale invariant, a necessary condition is that the contribution from this vertex be scale invariant,
barring miraculous cancellations. This implies that the scale factor must be growing as
\be
a \sim \frac{1}{(-y)}~,
\label{yinf}
\ee
which corresponds to an effective de Sitter geometry. Remarkably, simply demanding scale invariance of the two- and three-point correlation functions,
without any consideration of the independent dynamics of $a$ and $c_s$, has led us to focus on backgrounds that are effectively de Sitter, albeit in terms
of the $y$ variable. Thus the question becomes --- is it possible to have inflation without inflation? By this, we mean, is there an evolution where the
modes see an effective de Sitter space in terms of the $y$ variable but for which the true geometry is far from de Sitter? Unfortunately, the answer appears
to be no, as we now argue.

With $a(y)\sim 1/(-y)$,~(\ref{inv}) immediately implies that $\epsilon/c_s =\gamma$, where $\gamma$ is an arbitrary (positive) constant.
This is all we need to solve~(\ref{diffeq}), with the result
\be
c_s(y) = \frac{-1}{\gamma\log\left(y/\bar y\right)}~;~~~~~~~~~~~~~~~~~\epsilon(y) = \frac{-1}{\log\left(y/\bar y\right)}~;
\label{zero}
\ee
where $0 \leq \lvert y\rvert \leq \lvert\bar y\rvert$. Both $\epsilon$ and $c_s$ start out infinite and decrease rapidly to zero.  By construction, this solution is scale invariant at the two--point level and the aforementioned three--point vertex is also scale invariant. At first sight, we might expect that this solution is far from de Sitter because $\epsilon \gg 1$ initially. However, because $\epsilon$ is decreasing so rapidly, by the time $|y| < e^{-1}|\bar y|$; $\epsilon$ is already less than unity, indicating an inflationary spacetime. As such, this solution is only a small deformation away from the de Sitter geometry, specifically only about one e--fold of evolution is non--inflationary. 

It is worth noting that while the particular vertex~(\ref{S3particular}) we have considered yields a time-independent (and therefore scale invariant) non-gaussian contribution
once~(\ref{yinf}) is imposed, some of the other vertices vary logarithmically with $y$, such as the $\zeta'\partial\zeta\frac{\partial\zeta'}{\nabla^2}$ vertex. But this slow growth
as a function of scale is of course acceptable from the viewpoint of avoiding large non--gaussianities and the breakdown of perturbation theory. 

The analysis thus far strongly suggests that if we want scale invariance at the level of the three-point function, we are forced to consider an evolution that is very close to de Sitter --- that is, the cosmological background must inflate. The intuition for this statement is that every term in the cubic action is multiplied by powers of $\epsilon$. In the case of slow-roll inflation, we take $\epsilon \ll 1$ and therefore get a nearly gaussian spectrum. Note that there is another factor common to every term, namely $1/c_s^2$. One might therefore suspect that a nearly gaussian spectrum could also result from taking $c_s \gg 1$,
even with a strongly scale dependent $f_{\rm NL}$. This will be treated separately in Section~\ref{loophole}, where we will show that other considerations should greatly constrain this case as well.

One may wonder whether the assumption of scale invariant three-point function is not too stringent. A possible loophole would be to have non-gaussianities peak in the IR, where the onset of strong coupling
would lie on unobservably large scales. However it is clear from the above analysis that this would require a scale factor that expands more rapidly than $a\sim1/y$, which is super--inflationary in terms of the $y$ variable. We have studied this case by generalizing the above analysis to power-law dependencies for $f_{\rm NL}$, and found that inflation (in terms of cosmic time) remains the only solution.

\section{Superluminal $c_s$ Cosmologies}
\label{loophole}
Given the above arguments, it seems as though it is impossible to have small non-gaussianities over a broad range of modes without invoking a period of inflation. However, there is an illuminating exceptional case which we now treat separately. As we noted in passing earlier, every term in the cubic action has a factor of $1/c_s^2$ multiplying it. In analogy with slow--roll inflation, we may imagine controlling the size of the three-point function by making this factor very small, corresponding to superluminal $c_s$. In principle, this should allow for the production of modes with an extremely gaussian spectrum over a broad range in $k$--space~\cite{bimetric2}. In practice, however, we will find that only a finite range of scale invariant and gaussian modes can be generated by such mechanisms.

Superluminal scenarios have been proposed recently~\cite{magueijo}$-$\cite{tachyacoustic}, building on previous work on decaying sound speed cosmologies~\cite{vsl}$-$\cite{piao}.
These models take advantage of having an extremely large speed of sound at early times which suppresses non--gaussianities on large scales.  By virtue of the fact that the sound horizon at early times is much larger than the co--moving Hubble radius, the horizon problem is addressed. The large sound speed decays as time goes on, causing the co--moving sound horizon to shrink in time. See Fig.~\ref{horizons}. In the same manner that a shrinking Hubble horizon causes modes to freeze out in inflation, modes freeze out when they exit the sound horizon in this case. Additionally, the flatness and horizon problems are effectively decoupled in this scenario.
Of course, these models manifestly rely on superluminality, which, as mentioned before, makes them unsuitable for UV completion within a theory that is local in the usual sense~\cite{nima}. 

Current models in the literature~\cite{magueijo,bimetric,tachyacoustic} are cast in terms of constant $\epsilon$ cosmologies mentioned earlier and studied extensively in \cite{piazza}. In these scenarios, we assume scaling solutions for $a$ and $c_s$ given by~(\ref{acspowerlaw}):
\begin{figure}
   \centering
   \includegraphics[width=2.5in]{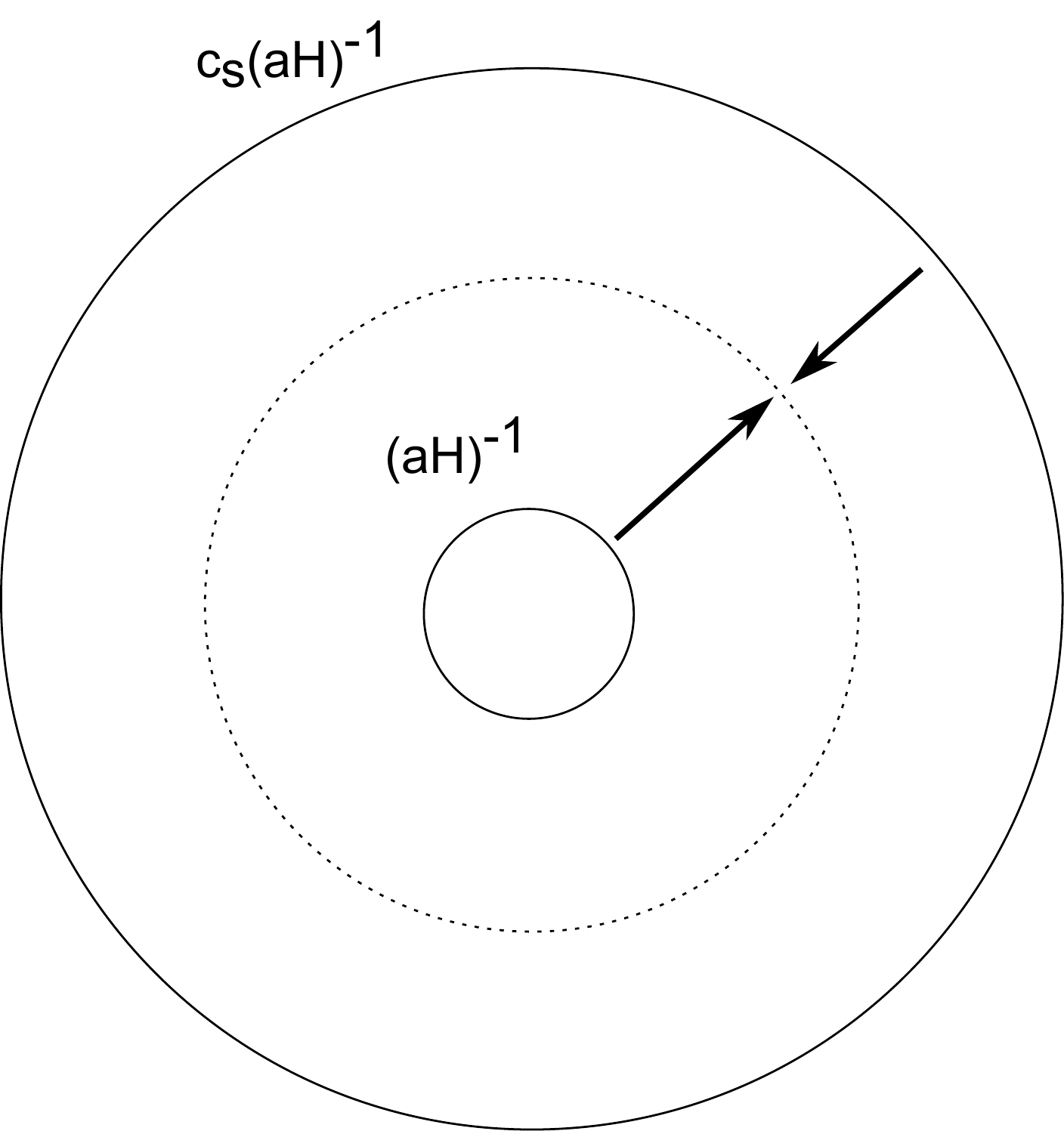}
   \caption{The initial setup with the sound horizon far outside the Hubble horizon. As the $y$ variable evolves, the comoving sound horizon decreases and the comoving Hubble horizon grows until the two meet at the coincidence scale denoted by the dashed line.}
   \label{horizons}
\end{figure}
\be
a=\left(\frac{y}{y_1}\right)^{\frac{1}{\epsilon+\epsilon_s-1}}~;~~~~~~~~~~~~~~~~c_s=\left(\frac{y}{y_1}\right)^{\frac{\epsilon_s}{\epsilon+\epsilon_s-1}}~,
\label{acsgg1}
\ee
where $y_1$ is the time when $c_s=1$. Using the freedom to rescale spatial coordinates, we have set $a=1$ at this time also. 
The Hubble parameter is given by 
\be
H=\frac{1}{y_1(\epsilon+\epsilon_s-1)}\left(\frac{y}{y_1}\right)^{\frac{-\epsilon}{\epsilon+\epsilon_s-1}}~.
\label{Hcsgg1}
\ee
The contribution to $f_{\rm NL}$ given by~(\ref{fNLeps3}) scales as
\be
f_{\rm NL}^{\epsilon^3} \sim \left(\frac{\epsilon}{c_s}\right)^2 \sim \frac{1}{y^4}\bigg\vert_{k|y| = 1} \sim k^4 \,,
\label{fNLcsgg1}
\ee
which thus peaks on small scales. Despite this strong scale dependence, the range of modes accessible to observations can be highly gaussian by having $c_s \gg 1$,
which amounts to making the coefficient in~(\ref{fNLcsgg1}) very small.

First note that  the Hubble horizon at mode freeze-out and the amplitude of the power spectrum have a simple relationship. 
Indeed, the power spectrum in the constant $\epsilon$ case is given by~\cite{piazza}
\be
P_{\zeta}=\frac{(\epsilon+\epsilon_s-1)^2}{8\pi^2\epsilon}\frac{H^2}{c_sM_{\rm Pl}^2}\bigg\vert_{k|y| = 1} =\frac{1}{8\pi^2\epsilon~y_1^2}\left(ky_1\right)^\frac{2\epsilon+\epsilon_s}{\epsilon+\epsilon_s-1}~,
\label{Pcsgg1}
\ee
which confirms the statement made earlier that exact scale invariance requires $\epsilon_s = -2\epsilon$. We will henceforth assume that $\epsilon_s = -2\epsilon$. 
It then follows from~(\ref{Pcsgg1}) that the Hubble parameter and the power spectrum are related by
\be
H^2 = M_{\rm Pl}^2 \frac{8\pi^2\epsilon}{(1+\epsilon)^2} c_s P_\zeta~,
\label{H2interm}
\ee
where everything is evaluated at freeze-out. Staring at this expression, it is immediately clear that in the limit $c_s \gg 1$ a tension arises between matching
the observed normalization of $P_\zeta\sim10^{-10}$ and keeping $H < M_{\rm Pl}$.

To be completely general, suppose that the phase of decaying sound speed lasts from some initial time, $y_{\rm initial}$,
to some final time $y_{\rm final}$. Without loss of generality, we will assume that $y_{\rm final} \leq y_1$ --- there is no gain in having
the decaying $c_s$ phase lasts any longer, since the modes generated when $y > y_1$ will never be outside the Hubble horizon.
Thus the range $N$ of scale invariant modes is given by $y_{\rm initial} = N y_{\rm final}$. Substituting this into~(\ref{Hcsgg1}),
we find that the ratio of the Hubble parameter at the beginning and end of the decaying $c_s$ phase satisfies
\be
\frac{H_{\rm initial}}{H_{\rm final}} = \left(\frac{y_{\rm initial}}{y_{\rm final}}\right)^\frac{\epsilon}{1+\epsilon} = N^\frac{\epsilon}{1+\epsilon}~.
\label{Hratio}
\ee
Meanwhile, from~(\ref{H2interm}) we have 
\be
\frac{H_{\rm final}}{M_{\rm Pl}} \simeq 10^{-5} \frac{\sqrt{8\pi^2\epsilon}}{1+\epsilon}\sqrt{c_s\rvert_{\rm final}}~,
\ee
where we have used the observational constraint $P_\zeta\sim 10^{-10}$. We therefore obtain
\be
\frac{H_{\rm intial}}{M_{\rm Pl}} \simeq 10^{-5} \frac{\sqrt{8\pi^2\epsilon}}{1+\epsilon}\sqrt{c_s\rvert_{\rm final}}N^\frac{\epsilon}{1+\epsilon}~.
\ee

For consistency of classical gravity, we demand that the Hubble parameter be sub--Planckian at the onset the decaying sound speed phase,
$H_{\rm initial} \lesssim M_{\rm Pl}$. This gives the following upper bound on $N$:
\be
N \lesssim \left(\frac{10^5}{\sqrt{8\pi^2}} \frac{1+\epsilon}{\sqrt{\epsilon}\sqrt{c_s\rvert_{\rm final}}} \right)^\frac{1+\epsilon}{\epsilon}~.
\label{bound1}
\ee
Remarkably, this bound on the range of scale invariant modes follows merely by requiring the correct amplitude for perturbations
and that the physics be sub--Planckian. So far we have said nothing about strong coupling issues. It turns out that the bound
can be made more precise by considering non-gaussianities. 

As before, consider the contribution to $f_{\rm NL}$ given by~(\ref{fNLcsgg1}). Demanding that this be consistent
with perturbation theory requires
\be
f_{\rm NL}^{\epsilon^3} \zeta\sim \left(\frac{\epsilon}{c_s\rvert_{\rm int}}\right)^2\cdot 10^{-5} \lesssim 1~,
\ee
which implies the lower bound $c_s\rvert_{\rm int} \,\gsim\, 10^{-\frac{5}{2}}\cdot\epsilon$. Inserting this into~(\ref{bound1}),
the bound on $N$ becomes
\be
N \lesssim\left[\frac{10^\frac{25}{4}}{\sqrt{8\pi^2}}\left(\frac{1+\epsilon}{\epsilon}\right)\right]^\frac{1+\epsilon}{\epsilon}~.
\label{bound2}
\ee
Of course we are interested in the case $\epsilon \geq 1$, for otherwise the universe would be inflating. The bound~(\ref{bound2}) is weakest for $\epsilon \simeq 1$,
which is sensible, since in this regime $c_s$ has to be just barely superluminal in order to suppress non--gaussianities. In this case,~(\ref{bound2})
enforces $N\sim 10^{12}\sim e^{28}$ for the range of scale invariant modes in $k$--space. While this is sufficient to account for microwave background
and large scale structure observations, it does represent a relatively narrow range within which the observable window must lie.

To summarize, it appears that strong coupling considerations also greatly constrain scenarios based on superluminality. The bound on the allowed range of
scale invariant modes is weaker than in the canonical case ($N \lesssim 10^{12}$ versus $ \lesssim 10^{5}$, respectively), but is nevertheless restrictive.
The scenario must somehow explain the coincidence of why the observational window happens to fall within the limited scale invariant range. Although here we have focused on the particular realization (\ref{acsgg1}), we expect that similar arguments will bound mode production in other realizations of the limit $c_s \gg 1$.

\section{Concluding Remarks}

A fundamental objective of early universe cosmology is to explain the origin of the observed spectrum of density perturbations.
The primordial perturbations inferred from observations are nearly scale invariant and highly gaussian. Any mechanism that purports to explain
their origin should be capable of generating perturbations with the desired properties over many decades in $k$--space.

The present paper concludes a body of work, initiated in~\cite{godfrey,baumann}, aimed at identifying {\it all} possible cosmologies
capable of generating scale invariant and gaussian perturbations. The analysis assumes a single degree of freedom (purely adiabatic
perturbations), and a background which represents a dynamical attractor. In the case of $c_s = 1$, it was shown in~\cite{godfrey,baumann} that only
inflation can generate suitable perturbations over many decades in $k$-space. Two other phases, corresponding to adiabatic ekpyrosis, can produce at
most $10^5$ modes (or 12 e-folds) of scale invariant and gaussian modes, before hitting strong coupling. Baumann {\it et al.}~\cite{baumann}
also considered a time-dependent sound speed, but their analysis was restricted to sub--luminal $c_s$ and nearly constant $\epsilon$. 

In this paper, we have generalized the analysis by allowing for {\it arbitrary} time-dependent $\epsilon(t)$ and $c_s(t)$,
with the only restriction that the null energy condition be satisfied: $\epsilon \geq 0$. At the two-point level, any cosmological background can in principle
yield a scale invariant spectrum, provided that the sound speed evolves in a suitable way. To avoid strong coupling we have argued that the three-point function should generically also be nearly scale invariant. By demanding that a particular contribution to $f_{\rm NL}$ be scale invariant, we have found that
inflation remains the unique solution. Consequently, non--inflationary solutions have strongly scale--dependent non--gaussianities, and therefore 
the range of modes consistent with observations is subject to a bound similar to the canonical case.

One exceptional case is the limit of superluminal sound speed, which has attracted considerable attention recently~\cite{magueijo}$-$\cite{tachyacoustic}. For concreteness, we have focused on superluminal scenarios where both $\epsilon$ and $\epsilon_s$ are constant in time.
In this case, non-gaussianities are strongly scale dependent, but strong coupling is na\"ively avoided by making $c_s \gg 1$. While this does alleviate the
constraints, we have shown that the combined requirements of sub--Planckian Hubble parameter and controlled non-gaussianities imply a bound of at most
$10^{12}$ modes (or 28 e-folds) with nearly scale invariant and gaussian statistics. This bound is weaker than in the canonical case, but is nevertheless
quite restrictive. Although we have not considered it in detail here, we expect arguments similar to those in Section \ref{loophole} to bound mode production in more general superluminal scenarios.

We conclude that, under the assumptions of single degree freedom, attractor background and adiabatic vacuum initial conditions, inflation is the only mechanism capable of generating
suitable perturbations over a wide range of scales. Alternative mechanisms must inevitably rely either on an instability of the background, non-vacuum initial state, 
and/or multiple degrees of freedom.

\textit{Note added:} As this paper was being completed, Reference \cite{kinney} appeared, which reaches conclusions similar to ours using different arguments. Specifically, the authors find that that in order to produce more than 7 e--folds of scale invariant modes in a monotonically expanding universe without inflation --- barring super--Planckian energy densities --- a superluminal speed of sound must be invoked. However, this result relies on the assumption of an expanding background, while our arguments do not require this constraint.

\textit{Acknowledgments:} It is our pleasure to thank Bryan Chen, Kurt Hinterbichler, Mitchell Lerner, Godfrey Miller and Leonardo Senatore for helpful discussions. This work was supported in part by funds from the University of Pennsylvania, NSF grant PHY-0930521, and the Alfred P. Sloan Foundation.

\end{document}